\documentclass{aa}

\usepackage{graphics}

\def\spose#1{\hbox to 0pt{#1\hss}}
\def\gsim{\mathrel{\spose{\lower 3pt\hbox{$\mathchar"218$}}
          \raise 2.0pt\hbox{$\mathchar"13E$}}}
\def\lsim{\mathrel{\spose{\lower 3pt\hbox{$\mathchar"218$}}
          \raise 2.0pt\hbox{$\mathchar"13C$}}}

\begin{document}

\title{The lag and duration-luminosity relations \\
of gamma-ray burst pulses}

\author{S. Boci\inst{1}, M. Hafizi\inst{1} and R. Mochkovitch\inst{2} }
\offprints{R. Mochkovitch}
\institute{$^1$Tirana University, Faculty of Natural Sciences, Tirana, 
Albania\\
           $^2$Institut d'Astrophysique de Paris, UMR 7095  
Universit\'e Pierre et Marie Curie-Paris 6 -- CNRS,
98 bis boulevard Arago, 75014 Paris, France}
\titlerunning{The time lag-luminosity relation of GRBs}

\abstract
{Relations linking the temporal or/and spectral properties of the prompt
emission of gamma-ray bursts (hereafter GRBs) to the absolute
luminosity are of great importance as they both constrain the 
radiation mechanisms and represent potential
distance indicators. Here we discuss two such relations: the
lag-luminosity relation and the newly discovered duration-luminosity
relation of GRB pulses.}
{We aim to extend our previous work on the origin of spectral lags, using the
duration-luminosity relation recently discovered by Hakkila et al.
to connect lags and luminosity. We also present a way to test this
relation which has originally been established with a limited sample of only
12 pulses.}
{We relate lags to the spectral evolution and shape of the pulses with
a linear expansion of the pulse properties around maximum. We then
couple this first result to the
duration-luminosity relation to obtain the lag - luminosity and lag -
duration 
relations. We finally use a Monte-Carlo method to generate a population
of synthetic GRB pulses which is then used to check the validity of the
duration-luminosity relation.   }
{Our theoretical results for the lag and duration-luminosity relations 
are in good agreement with the data. They are rather insensitive to the
assumptions regarding the burst spectral parameters. Our Monte Carlo
analysis of a population of synthetic pulses confirms that the 
duration-luminosity
relation must be satisfied to reproduce the observational duration --
peak flux diagram of BATSE GRB pulses.  }
{The newly discovered duration-luminosity relation 
offers the possibility to link all three quantities: lag, duration and
luminosity of GRB pulses in a consistent way. Some evidence for its
validity 
have been presented but its origin is not easy to explain in the context
of the internal shock model.}

\titlerunning{Lag and duration-luminosity relations}
\maketitle 
   
\section{Introduction}
The prompt emission of gamma-ray bursts is
characterized by the diversity of the observed temporal profiles. Some bursts
show a simple shape with a fast rise followed by a slower decay while
others have a complex structure with a succession of pulses which can be
overlapping or separated by intervals with almost no
emission. Conversely the spectra are more uniform, generally well fitted
by two smoothly connected power laws (the so-called Band spectrum; Band 
et al., 1993). Many
studies have tried to link the temporal and spectral properties of
bursts with the objective to gain insight into the physical processes
governing the prompt emission. Already before and during the BATSE era 
several relations
between hardness and duration (Kouveliotou et al, 1993), intensity 
(Golenetskii et al., 1983) 
and fluence (Liang \& Kargatis, 1996) were
found. Following the 
discovery of the afterglows and the measure of the first redshifts,
intrinsic quantities such as the luminosity or the total radiated energy
became accessible and new
relations appeared: the Amati (Amati et
al, 2002) and Ghirlanda (Ghirlanda et al., 2004) relations 
between the peak energy of the
global 
spectrum and the energy
release in gamma-rays (assuming isotropic emission in the Amati
relation and corrected for beaming in the Ghirlanda relation), the
luminosity-variability relation (Reichart et al., 2001) 
illustrating the tendency of luminous
bursts to be more variable and the lag-luminosity relation (hereafter
LLR)
discovered by
Norris et al. (2000). Spectral lags are a way to quantify the changes
in the burst profiles observed in different energy bands. When viewed  
at high energy, pulses are narrower and peak earlier. Norris et al.
(2000) cross-correlated the profiles between BATSE bands 1 and 3 and
found that the resulting lags were decreasing with increasing burst peak
luminosity. As for the other relations between luminosity and spectral
or temporal properties, the LLR offers clues to
the physics of the prompt emission but also provides a potential method
to evaluate GRB distances from observations at high energy only.

Spectral lags are a direct consequence of the burst spectral evolution
since a fixed, constant spectrum, would lead to proportional
profiles in all energy bands. In a first paper (Hafizi \& Mochkovitch,
2007) we computed spectral lags of pulses, defined as the
time interval between pulse maximum in two different bands. We obtained
an explicit expression for the lags and assuming the validity of an
``Amati-like'' relation between luminosity and the value of $E_{\rm p}$
at pulse maximum we were able to connect lags and luminosity. 

Hakkila et al. (2008) have recently reconsidered the LLR and obtained a
new relation which applies to individual pulses while the original Norris et al.
(2000)
LLR considered the burst as a whole. Moreover, they also found a correlation 
between pulse duration and luminosity (hereafter DLR). These results offer 
the possibility to directly test and extend our previous work  
(Hafizi \& Mochkovitch, 2007). We start in Sect.2 by comparing to
observations our 
theoretical results for the LLR. Since
they rely
on the validity of the DLR we propose to test it in Sect.3 by comparing a
synthetic population of GRB pulses to the
observed peak flux -- duration diagram of a sample of pulses
collected by Hakkila \& Cumbee (2009). We discuss our results in Sect.4
and Sect.5 is the conclusion.

\section{The lag-luminosity relation}
\subsection{Theoretical interpretation}
In paper I (Hafizi \& Mochkovitch, 2007) 
we presented a simple analytical model 
to calculate spectral
lags. We recognized that lags were better defined using individual
pulses rather than the whole burst profile. Pulses in the same burst can
have different lags and Hakkila et al. (2008) have shown that the global lag 
represents some average where the brightest pulse (which generally has
the shortest lag) makes the dominant contribution. Looking for
correlation between lag and luminosity it is therefore preferable to
consider each pulse separately. Hakkila et al. (2008) obtained
\begin{equation}
L=6.1\,10^{52}\ (\Delta t_{13}/0.01\ {\rm s})^{-0.62}\ \ \ {\rm erg.s}^{-1}
\end{equation}
where $L$ is the peak pulse luminosity and $\Delta t_{13}$ the spectral
lag between BATSE bands 1 and 3. This expression is believed to be more
directly linked to the physics of the prompt emission 
than the original one found by Norris et al. (2000)
\begin{equation}
 L=1.3\,10^{53}\ (\Delta t_{13}/0.01\ {\rm s})^{-1.15}\ \ \ {\rm erg.s}^{-1}
\end{equation}
where lags were computed by cross-correlation of the full burst profile
between the two bands. For individual pulses, spectral lags are more
easily estimated from the time difference between the peaks. 
These ``pulse peak lags'' generally agree with those obtained by
cross-correlation and have been used by Hakkila et al. (2008) to get
Eq.(1).

In our theoretical analysis (Hafizi \& Mochkovitch, 2007) we 
calculated pulse peak lags from a linear expansion of the pulse shape
and spectral properties around the maximum in BATSE band 1 (20 - 50
keV). 
Our result directly relates the
lag to spectral evolution in a very transparent way. We get
\begin{equation}
{\Delta t_{13}\over t_{\rm p}}= {f_{13,E_{\rm p}}\,{\dot e}_{\rm p}+
f_{13,\alpha}\,{\dot a}+f_{13,\beta}\,{\dot b}\over C_1}
\end{equation}
where $t_{\rm p}$ is the pulse duration. The other terms are defined in
the following way:
\begin{equation}
f_{13,X}={\partial Log {\cal F}_{13}\over \partial Log X}{\Big |}_{t_1}
\ \ {\rm with}\ \ {\cal F}_{13}={\int_{100/E_{\rm p}}^{300/E_{\rm p}} 
{\cal B}_{\alpha\beta}(x)\,dx
\over \int_{20/E_{\rm p}}^{50/E_{\rm p}} {\cal B}_{\alpha\beta}(x)\,dx}\ .
\end{equation}
Here $t_1$ is the time of pulse maximum in BATSE band 1 and 
${\cal B}_{\alpha\beta}(x)$ is the spectrum shape for which we assumed a
Band function (with $x=E/E_{\rm p}$, $E_{\rm p}$ being the peak energy of the spectrum and
$\alpha$ and $\beta$ the two spectral indices at low and high energy). 
The derivatives ${\dot e}_{\rm p}$, $\dot a$ and $\dot b$
are respectively given by
\begin{equation}
{\dot e}_{\rm p}={{\dot E_{\rm p}}\over E_{\rm p}}\,t_{\rm p},\ \  
{\dot a}={{\dot \alpha}\over \alpha}\,t_{\rm p},\ \  
{\dot b}={{\dot \beta}\over \beta}\,t_{\rm p}
\end{equation}
and are evaluated at $t=t_1$. Finally $C_1$ is a ``curvature
parameter'' for the pulse around maximum. We have
\begin{equation}
{C_1\over t_{\rm p}^2}={{\ddot N}_1(t_1)\over N_1(t_1)}\ \ \ \ \ \ (C_1<0)
\end{equation}
where $N_1(t)$ is the count rate in BATSE band 1.

The functions $f_{13,X}$ depend on the spectrum parameters $E_{\rm p}$, 
$\alpha$ and $\beta$ at pulse maximum while ${\dot e}_{\rm p}$, $\dot a$
and $\dot b$ represent their evolution. The two remaining quantities in 
Eq.(1), $t_{\rm p}$ and $C_1$, are fixed by the pulse shape. They show
that spiky pulses (large $|C_1|$) have shorter lags than broad pulses
(small $|C_1|$) for a given spectral evolution and pulse duration and that
short pulses are expected to have short lags, both effects in agreement
with observations (Norris \& Bonnell, 2006; Gehrels et al., 2006; 
Hakkila et al., 2007). The three functions $f_{13,E_{\rm p}}$,
$f_{13,\alpha}$ and $f_{13,\beta}$ have been represented in Fig.1 for 
$\alpha=-1$ and $\beta=-2.25$ which are the central values of the
distributions found by Preece et al. (2000) in their study of the spectral properties of 
bright BATSE bursts. Their behavior can be understood by noting that at
large (resp. small) $E_{\rm p}$ values 
${\cal F}(E_{\rm p},\alpha,\beta)$ depends on $\alpha$ (resp. $\beta$)
only. Therefore 
$f_{13,E_{\rm p}}=\partial Log {\cal F}_{13}/ \partial Log E_{\rm p}$ 
mostly contribute at intermediate $E_{\rm p}$ (between BATSE bands 1 and
3) while $f_{13,\alpha}$ (resp. $f_{13,\beta}$) dominates at large
(resp. small) $E_{\rm p}$. The resulting ratio $\Delta t_{13}/t_{\rm p}$
has been plotted in Fig.2 as a function of $E_{\rm p}$ for different
values of ${\dot e}_{\rm p}$, $\dot a$, $\dot b$ and $|C_1|=10$. 
\begin{figure*}\center{
\resizebox{.6\hsize}{!}{\includegraphics{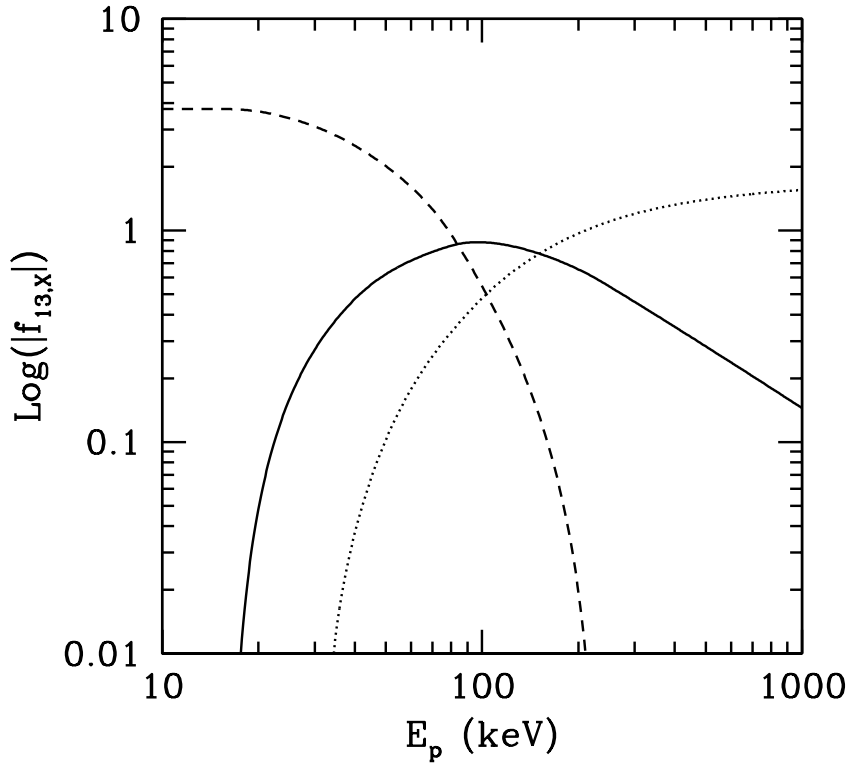}}}
\caption{Plot of the three functions $f_{13,E_{\rm p}}$ (full line),
$-f_{13,\alpha}$ (dotted line) and $-f_{13,\beta}$ (dashed line) for
$\alpha=-1$ and $\beta=-2.25$.} 
\end{figure*} 
\begin{figure*}\center{
\resizebox{.6\hsize}{!}{\includegraphics{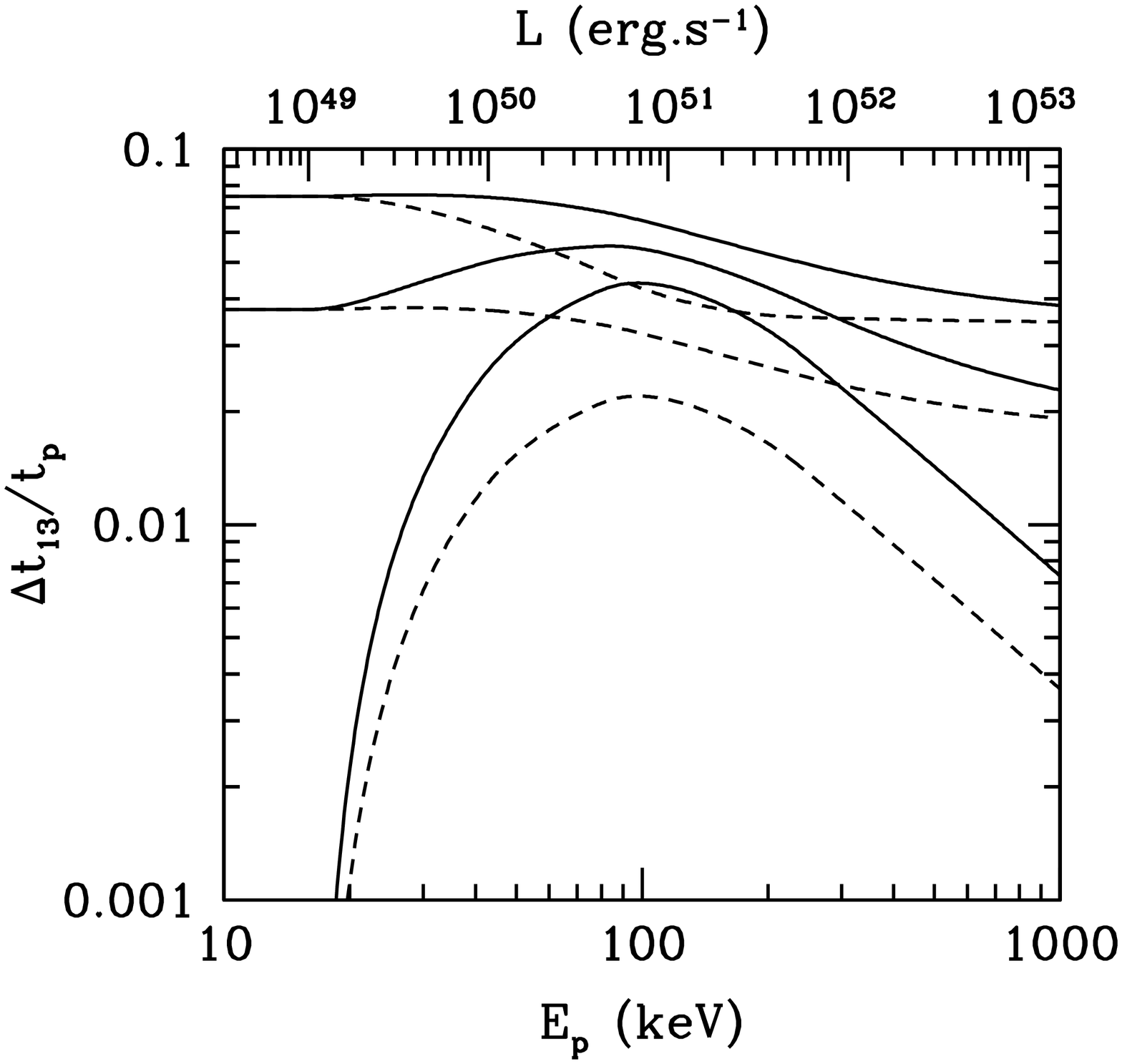}}}
\caption{Ratio of spectral lag over pulse duration as a function of the
 peak energy at pulse maximum (lower scale) and of the peak luminosity
 (upper scale) assuming the validity of the Amati-like relation (Eq.(7)). The
 full lines correspond to ${\dot e}_{\rm p}=-0.5$ and from top to bottom
 respectively to $\dot a=\dot b=0.2$, $0.1$ and $0$. The dashed lines
 have the same $\dot a$ and $\dot b$ but ${\dot e}_{\rm p}=-0.25$. The 
 adopted value of the curvature parameter is $|C_1|=10$.    
}
\end{figure*}   

This value of the curvature parameter as been adopted as representative
of a ``typical pulse''. In any case, the results for a different $|C_1|$
are easily obtained by rescaling $\Delta t_{13}/t_{\rm p}$ by a factor 
$10/|C_1|$.
As the maximum of $E_{\rm p}$ generally precedes that of the count rate
in most pulses we have ${\dot e}_{\rm p}< 0$. Similarly, the decrease of
the spectral indices (spectral softening)
begins before pulse maximum implying that $\dot a>0$
and $\dot b>0$ (since $\alpha$ and $\beta$ are negative). 

To link spectral lags and luminosity Hafizi \&
Mochkovitch (2007) have moreover assumed an ``Amati-like relation'' 
between $E_{\rm p}$ and
the pulse peak luminosity of the form
\begin{equation}
E_{\rm p}=380\left(L\over 1.6\,10^{52}\; {\rm erg.s}^{-1}\right)^{0.43}\
 {\rm keV}.
\end{equation}
This relation, proposed by Ghirlanda et al. (2005), is expected to be
valid at any time contrary to the original Amati relation (Amati et al.,
2002) which
applies to the burst as a whole. Using Eq.(3) and (7) it becomes
possible to represent $\Delta t_{13}/t_{\rm p}$ as a function of
luminosity. The results have also been plotted in Fig.2. It can
be seen that, as long as $\dot a$ and $\dot b$ are not too close to 0,
$\Delta t_{13}/t_{\rm p}$ only weakly depends on the luminosity. 
It is only if the spectral
evolution is limited to a decrease of the peak energy (the spectral
indices staying constant) that $\Delta t_{13}/t_{\rm p}$ can reach very
low values at both high and low luminosities. Since observed
bursts show a simultaneous spectral evolution in $E_{\rm p}$ and the
spectral indices we expect that $\Delta t_{13}/t_{\rm p}$ will not
change much from pulse to pulse in agreement with observations 
(Hakkila et al., 2008).

A roughly constant value of $\Delta t_{13}/t_{\rm p}$
however raises a problem which was already mentioned in paper I. If
pulses indeed satisfy a lag-luminosity relation with bright pulses
having very short lags, some additional
parameter has to be correlated to the luminosity. In paper I we tentatively
proposed that pulse curvature could be such a parameter, luminous pulses
being spikier and less luminous ones broader. But the recent discovery
by Hakkila et al. (2008) of a possible correlation between pulse duration
and peak luminosity offers a new perspective which can naturally account for
the LLR.
\begin{figure*}\center{
\resizebox{.6\hsize}{!}{\includegraphics{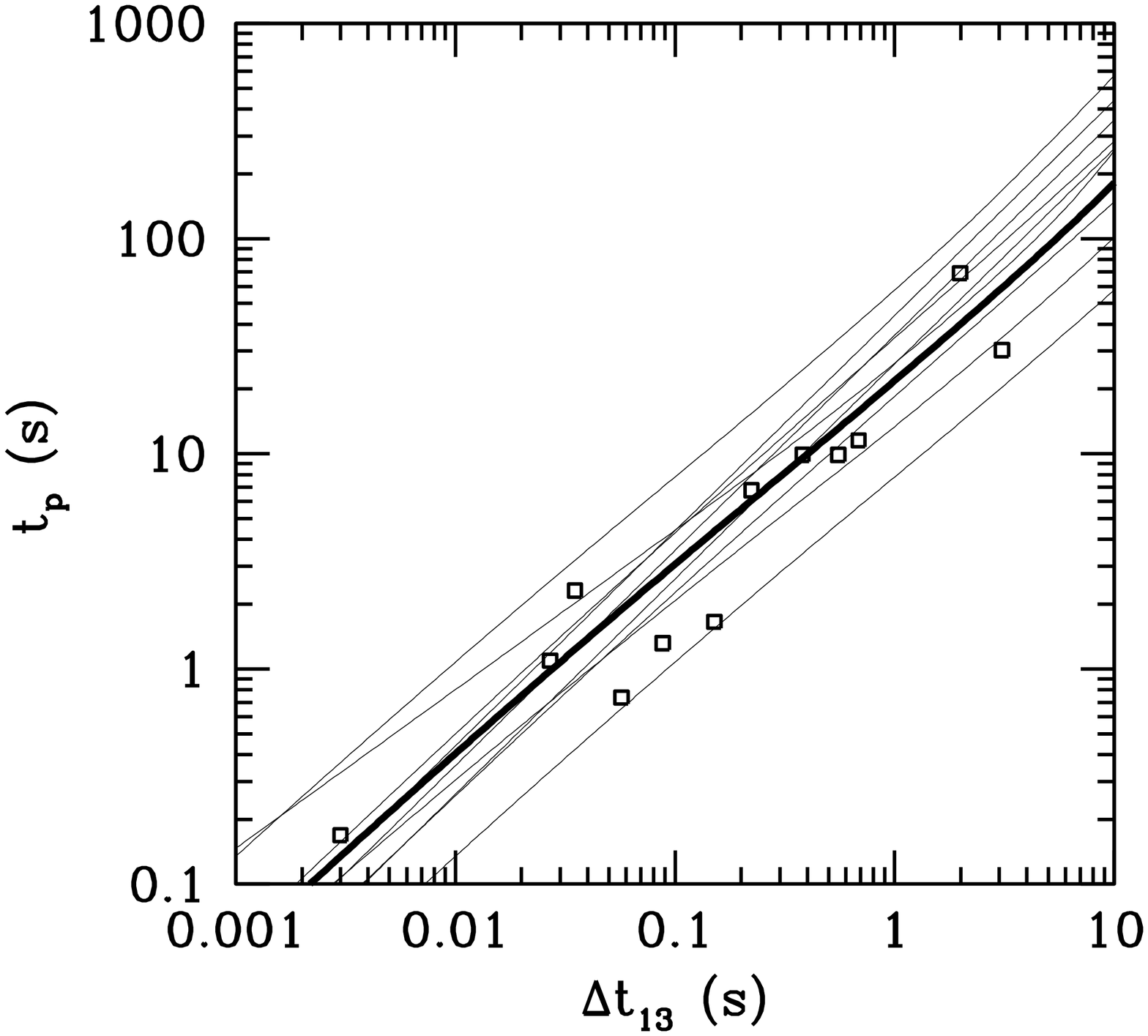}}} 
\caption{Plot of the pulse duration as a function of spectral lag. 
The full thick
line represents our reference model which adopts the Amati-like relation 
(Eq.(7)) 
and  $\alpha=-1$, $\beta=-2.25$, 
$\dot E=-0.5$, $\dot a=\dot b=0.1$ and $|C_1|=10$. The thin lines 
correspond to nine other cases with different choices of the parameters
(see text for details). The squares are the data points of the sample
considered in Hakkila et al. (2008). }
\end{figure*}

\begin{figure*}\center{
\resizebox{.6\hsize}{!}{\includegraphics{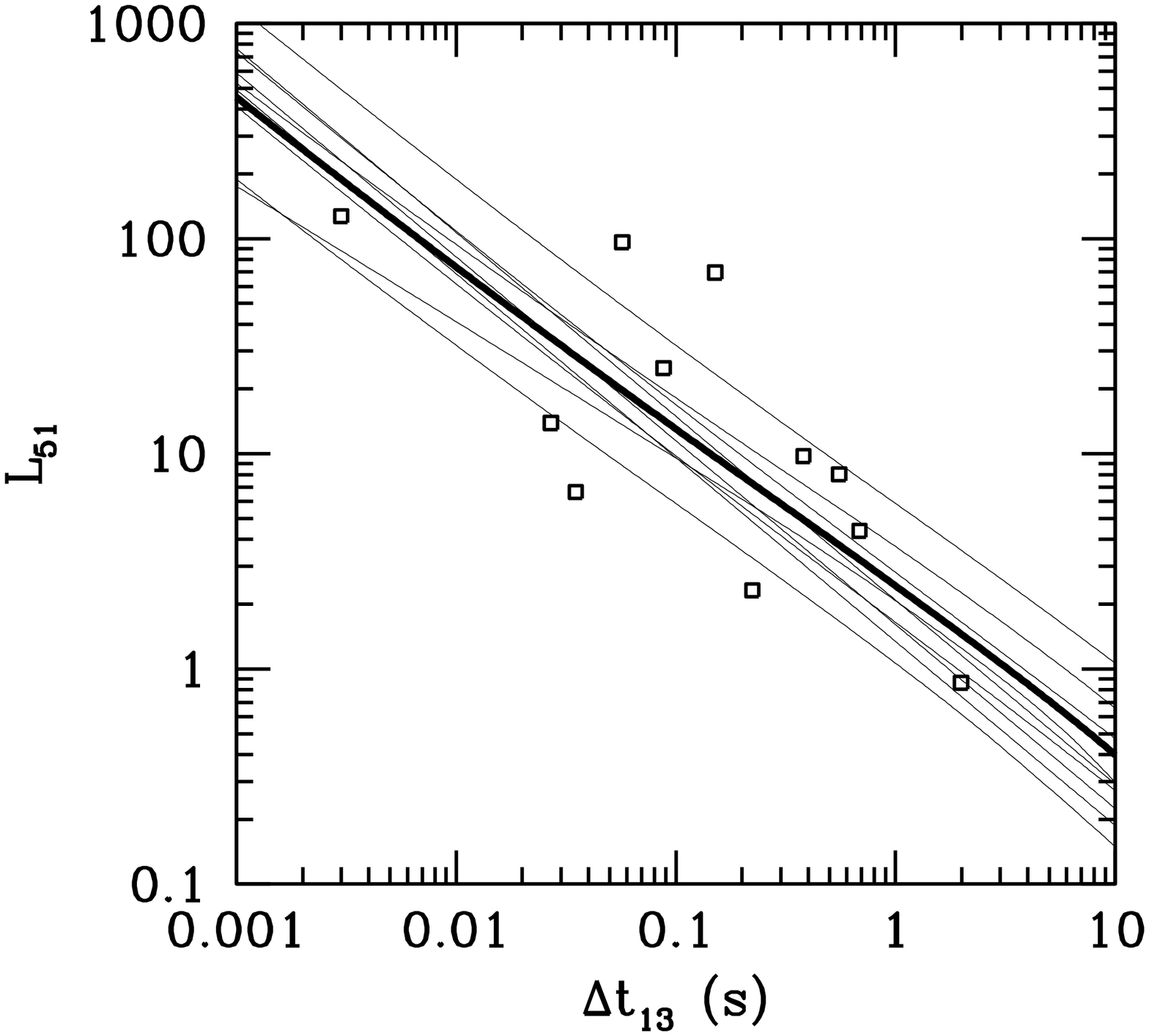}}}
\caption{Theoretical lag-luminosity relation for pulses compared to
the data collected by Hakkila et al. (2008). The thick line is
the reference case while the thin lines correspond to different model 
parameters (see Fig.3 and text for details).}  
\end{figure*}

\subsection{The duration-luminosity relation of GRB pulses}
Hakkila et al. (2008) re-analized the seven BATSE bursts with known
redshift considered by Norris et al. (2000) to establish the original LLR 
but they treated each pulse of these bursts separately. From the 12
selected pulses they obtained a new LLR (Eq.(1)) but also discovered an
even tighter relation between duration and luminosity
\begin{equation}
L=3.4\,10^{52}\; t_{\rm p}^{-0.85}\ \ \ \ {\rm erg.s}^{-1}\ .
\end{equation}
Coupling this to Eq.(3) then provides a very simple way to get the
LLR. An important difference with paper I is that we do not necessarily 
assume the validity of the ``Amati-like relation'' (Eq.(7)) to link
$E_{\rm p}$ to the luminosity. Even if $E_{\rm p}$ and $L$ are 
essentially 
independent quantities a LLR can still be obtained with the same  
value of the 
peak energy at all luminosities. In Fig.3 and 4 we present the results 
for both the $\Delta t_{13}$ -  $t_{\rm p}$ relation and the LLR. The 
full heavy line in each diagram corresponds to our reference case which
assumes the validity of the Amati-like relation and uses the following
values of the parameters:
$\alpha=-1$, $\beta=-2.25$, ${\dot e}_{\rm p}= -0.5$, $\dot a=\dot b=0.1$ and
$|C_1|=10$. 
In addition we plot with thin lines a few other cases: (1)
${\dot e}_{\rm p} =-1$ (other
parameters unchanged); (2) ${\dot e}_{\rm p} =-0.25$; 
(3) $\dot a=\dot b=0$ (other
parameters unchanged); (4) $\dot a=\dot b=0.2$; (5), (6) and (7) no
Amati
relation but constant  
$E_{\rm p}=250$, 500 and 1000 keV; (8) and (9) similar to 1) but with
$|C_1|=3$ and $30$.
All these lines define a narrow strip showing that both relations
remain fairly  robust even when the parameters are varied by large 
factors. 

We have also plotted in Fig.3 and 4 the data points for
the pulses belonging to the sample studied by Hakkila et al. (2008). It can be
seen that the agreement with our theoretical results is satisfactory.

\section{A test for the duration-luminosity relation}
\subsection{Method}
The results presented in the last section rely on the validity of the
duration-luminosity relation (DLR) for pulses. If confirmed, the DLR
would also
offer a new method to estimate GRB distances, simpler and
easier to use than the LLR (Hakkila, Fragile \& Giblin, 2009). 
Bursts with several pulses give the
possibility of multiple measures of the redshift, increasing the
resulting accuracy. Conversely the identical redshift for all pulses in
a given burst allows to test the DLR. Assuming a power-law of the form 
$L\propto t_{\rm p}^{-s}$ Hakkila, Fragile \& Giblin (2009) 
find $s=0.8 \pm 0.4$
for a sample of 53 multi-pulsed GRBs, which is consistent with the result
obtained from bursts with known redshift. 

We have performed an alternative and independent test of the DLR using a
synthetic population of GRB pulses for which we predict the resulting
observational duration -- peak photon flux 
($t_{\rm p}^{\rm obs}$ -- $P$) 
diagram which is then compared to real
data. The synthetic population is generated following a Monte-Carlo
method similar to the one described in Daigne, Rossi \& Mochkovitch
(2006): 
for each pulse we draw a
redshift $z$ and a peak luminosity $L$. We then either 
link $E_{\rm p}$ and $t_{\rm p}$ to the luminosity with Eq.(7) and (8)
or adopt 
log-normal distributions independent of $L$. We want to see
if the predicted $t_{\rm p}^{\rm obs}$ -- $P$ diagram is in better
agreement with the data when the
DLR is adopted.

Even if $t_{\rm p}$ and $L$ are uncorrelated we expect
a first trend purely due to cosmological effects as more distant
pulses are globally weaker and have longer durations. If an intrinsic
relation such as Eq.(8) is also satisfied the trend should be more
pronounced. The analysis by Hakkila \& Cumbee (2009) of a sample of 
pulses extracted from 106 long GRBs
gives $P\propto t_{\rm p}^{-0.27}$ which is 
shallower than Eq.(8). The observed population is however affected by
cosmological effects (time dilation and k-correction) on the pulse
duration 
(Norris, 2002) and by 
selection effects due to instrument threshold. We model 
these different effects to generate the simulated observational 
$t_{\rm p}^{\rm obs}$ - $P$ diagram from our synthetic GRB pulse population.
\subsection{The synthetic pulse population}
To get the distribution of 
the pulse parameters $z$, $L$, $E_{\rm p}$ and $t_{\rm p}$ in the
synthetic sample we make the 
following assumptions:

- Redshift $z$: as we only consider long GRBs which have massive
  progenitors, the GRB rate $R_{\rm GRB}(z)$ could a priori be expected to be
  proportional to the cosmic star formation rate $SFR(z)$. However
  recent studies (Daigne, Rossi \& Mochkovitch, 2006; Guetta \& Piran,
  2007; Kistler et al., 2008) 
have shown that at large $z$ the GRB rate still
  increases while the SFR probably decreases or remains constant. This
  suggests that stellar populations at large $z$ are more efficient in
  producing GRBs for reasons which are not well understood (reduced
  metallicity or/and IMF favoring massive stars). In this study we
  adopt a burst rate which follows SFR3 of Porciani \& Madau (2001). This SFR
  which keeps rising at large $z$ is not realistic as it would
  overproduce metals at
  early cosmic times but $R_{\rm GRB}\propto $ SFR3$(z)$ provides a good fit
  of the redshift distribution of Swift bursts. We then generate a table
  of $N$ ($N$ from $10^3$ to $10^6$) values of the redshift and
  corresponding luminosity distance $d_{\rm L}(z)$ with $z$
  being distributed as 
\begin{equation}
{dN\over dz}=N{{dV\over dz} {SFR3(z)\over 1+z}\over 
\int_0^{\infty}{dV\over dz} {SFR3(z)\over 1+z}}
\end{equation}
where $dV\over dz$ is the comoving volume element in the concordance
cosmology.

- Luminosity $L$: we adopt a power law luminosity function
  $\Phi(L)\propto L^{-\delta}$ between $L_{\rm min}$ and $L_{\rm max}$.
For the burst population it has been shown that a power law LF with 
$1.5<\delta<2$ can reproduce the $Log\,N$ - $Log\,P$ curve (Firmani et
  al., 2004; Daigne, Rossi \& Mochkovitch, 2006).  
We adopt the same range of values here and vary $L_{\rm min}$
and $L_{\rm max}$
respectively from $10^{50}$ to $10^{51}$ erg.s$^{-1}$ 
and from $10^{53}$ to $10^{54}$ erg.s$^{-1}$. 

- Spectral parameters: the peak energy is either obtained from the
  luminosity with the Amati-like relation or has a log-normal
  distribution of central value $E_{{\rm p},0}$ and width
  $\sigma_0$ . When the Amati-like relation is adopted we add a
  dispersion $\sigma_{\rm A}$ around Eq.(7). We draw the spectral
  indices $\alpha$ and $\beta$ in agreement with the distributions
  found by Preece et al. (2000) for bright BATSE bursts. In Daigne, Rossi \&
Mochkovitch (2006) the
  values of $\sigma_0$, $E_{{\rm p},0}$ and $\sigma_{\rm A}$ were
  adjusted to provide a good fit of the $E_{\rm p}$ distribution of
  bright BATSE
  bursts. We keep the same values as a starting point but we also vary 
  them since we are
  now considering individual pulses rather than
  the entire bursts. 

- Duration: To get the pulse duration we either assume the validity of
 the DLR (Eq.(8)) with a dispersion $\sigma_{\rm t_{\rm p}}$ or adopt a
 log-normal distribution of central value and dispersion adjusted to
reproduce the observed distribution of pulse duration in the Hakkila \&
 Cumbee (2009) sample. 
\\
 
For a given pulse, the observable quantities $P$ and 
$t_{\rm p}^{\rm obs}$ are then computed from the intrinsic parameters. 
For the peak photon flux we have, assuming a normalized Band spectrum
(i.e. $\int_0^{\infty}{\cal B}(x)=1$) 
\begin{equation}
P={L\over 4\pi\,d_{\rm L}^2\,E_{\rm p}^{\rm obs}}\int_{50/E_{\rm p}^{\rm obs}}
^{300/E_{\rm p}^{\rm obs}}{\cal B}(x)dx
\end{equation}
where  
$E_{\rm p}^{\rm obs}=E_{\rm p}/(1+z)$. To obtain the observed pulse duration
we take into account both time dilation and the approximate dependence of pulse
width 
with energy  $t_{\rm p}\propto E^{-0.4}$ (Norris et al., 1996) so that 
\begin{equation}
t_{\rm p}^{\rm obs}\simeq (1+z)^{0.6}\,t_{\rm p} \ .
\end{equation}
We adopt the threshold prescription from Band (2003) where 
the limiting photon flux is defined between 1 and 1000 keV and depends on the
observed peak energy. This finally allows us to construct 
the simulated observational 
$t_{\rm p}^{\rm obs}$ -- $P$ diagram for comparison to real data.
\subsection{Results}
We define a reference case which corresponds to the
following choice of the parameters: slope of the luminosity function 
$\delta=1.7$; $L_{\rm min}=10^{51}$ erg.s$^{-1}$; peak energy obtained
from the Amati-like relation (Eq.(7)) with an added dispersion of 0.3
dex. We compare in Fig.5 the resulting 
$t_{\rm p}^{\rm obs}$ - $P$ diagrams with and without the DLR. When the
DLR
is adopted we again assume a dispersion of 0.3 dex around Eq.(8).
A fit of the diagrams by a power-law $P\propto t_{\rm p}^{-s}$
respectively gives 
$s=0.27$ (with the DLR) and 
$s=0.09$ (without). In the first case, the agreement 
with the data of Hakkila \& Cumbee (2009) is excellent while the
correlation almost disappears in the second case. 

We then checked how these results are changed when we vary the model
parameters
and assumptions (see Table 1):

- Luminosity function: We list the power-law index $s$ of the 
$t_{\rm p}^{\rm obs}$ -- $P $ relation when we vary the lower and
upper
limits of the luminosity function $L_{\rm min}$ and $L_{\rm max}$
and its slope $\delta$. It can be seen that the results only weakly depend on  
$L_{\rm min}$ and $L_{\rm max}$ and are nearly unsensitive to $\delta$.

- Peak energy distribution: we have first replaced the Amati relation by a 
log-normal distribution of central value 
$E_{\rm p}=600$ keV and dispersion 0.3 dex which were the values 
adopted by Daigne, Rossi \& Mochkovitch (2006). Since these 
corresponded
to the whole burst spectra and not to individual pulses we have 
considered 
other possible $E_{\rm p}$ values. 
We find that the  power-law index of the 
$t_{\rm p}^{\rm obs}$ -- $P$ relation is practically independent of 
the adopted  
$E_{\rm p}$, especially when we assume the validity of DLR. 
However in this case
the $t_{\rm p}^{\rm obs}$ -- $P$ relation becomes somewhat steeper than 
the data (with $s=0.33$).

- Dispersion of the DLR: we have increased the dispersion of the DLR from
$\sigma_{\rm t_{\rm p}}=0.3$ dex to 1.5 dex. We observe that the power-law
index of the 
$t_{\rm p}^{\rm obs}$ -- $P$ relation evolves from its reference value of 0.27
to 0.09 which corresponds to the situation without the DLR. 
It appears that the dispersion of the DLR cannot exceed about
0.6 dex if we still want to fit the data.     

\begin{tabular}[t]{||c|c|c||}
\hline 
$L_{\rm min}$ (erg.s$^{-1}$) & $s$ (with DLR) 
& $s$ (without DLR)\\
\hline
$10^{51}$ & 0.27 & 0.091 \\
\hline
$10^{50.5}$ & 0.24 & 0.093 \\
\hline
$10^{50}$ & 0.22 & 0.096 \\
\hline\hline 
$L_{\rm max}$ (erg.s$^{-1}$) & 
& \\
\hline
$10^{53.5}$ & 0.24 & 0.093 \\
\hline
$10^{54}$ & 0.22 & 0.096 \\
\hline \hline
$\delta$ & & \\
\hline
2.0 & 0.27 & 0.081 \\
\hline
1.5 & 0.27 & 0.096 \\
\hline \hline
$E_{\rm p}$ (keV) & & \\
\hline
400 & 0.33 & 0.089 \\
\hline
600 & 0.33 & 0.088 \\
\hline
800 & 0.33 & 0.096 \\
\hline \hline
$\sigma_{\rm t}$ & & \\
\hline
0.3 & 0.27 & - \\
\hline
0.6 & 0.22 & - \\
\hline
1.0 & 0.15 & - \\
\hline
1.5 & 0.09 & - \\
\hline \hline
Threshold & & \\
\hline
:2 & 0.30 & 0.11 \\
\hline
x2 & 0.24 & 0.085 \\
\hline
\end{tabular}
\\
 
\noindent
Table 1: Slope $s$ of a power law fit ($P\propto t_{\rm p}^{-s}$)
of the $t_{\rm p}$ - $P$ diagram with
and without the assumption of the DLR (Eq.(8)) for pulses. In the six blocks
we respectively vary the lower (i) and upper (ii) limits of the pulse luminosity
function; 
(iii) the slope of the luminosity function; (iv) the central 
value of a log-normal distribution for $E_{\rm p}$; (v) the dispersion
of the DLR; (vi) the detection threshold.  The first row corresponds to
our reference case with $L_{\rm min}=10^{51}$ erg.s $^{-1}$,
$L_{\rm max}=10^{53}$ erg.s $^{-1}$,
$\delta=1.7$, $\sigma_{\rm t}=0.3$ dex. It also assumes the validity of the
Amati-like relation (Eq.(7)) with a dispersion of 0.3 dex and adopts the
threshold criterion for BATSE given by Band (2003). In each block only
one parameter is varied, the others keeping the values corresponding to
the reference case.
\\

Our study then indicates that an intrinsic correlation between pulse
duration and luminosity  
is necessary to reproduce the observed $t_{\rm p}^{\rm obs}$ -- $P$
diagram. This conclusion is not affected when we vary   
the model paramaters such as the luminosity function or the spectral 
properties of pulses.
It is also robust regarding changes in 
the adopted sensitivity: when we
decrease (resp. increase)    
the  
threshold by a factor of two, the $t_{\rm p}^{\rm obs}$ -- $P$
correlation becomes only slightly steeper (resp. shallower).

Nevertheless, a few words of caution should be expressed since
our analysis does not take into account possible additional
selection
effects which may not apply equally to pulses of different durations.
For example the data has been collected with a 
trigger criterion
applied to the full burst and not to individual
pulses. It therefore includes some pulses below the threshold, coming
from bursts which triggered at a brighter instant of the light curve. 
Also, the pulse selection and identification technique can fail when 
many pulses overlap, which is another source of selection       
effects, not easy to quantify.   
It is possible that these different biases contribute to produce
an effective
threshold with a limit in fluence in addition to the adopted
limit in peak flux. A limit in fluence may artificially generate a
trend
in the $P$ -- $t_{\rm p}^{\rm obs}$ diagram which could contribute to the
observed relation. 

\begin{figure*} 
\begin{center}
\resizebox{.9\hsize}{!}{\includegraphics{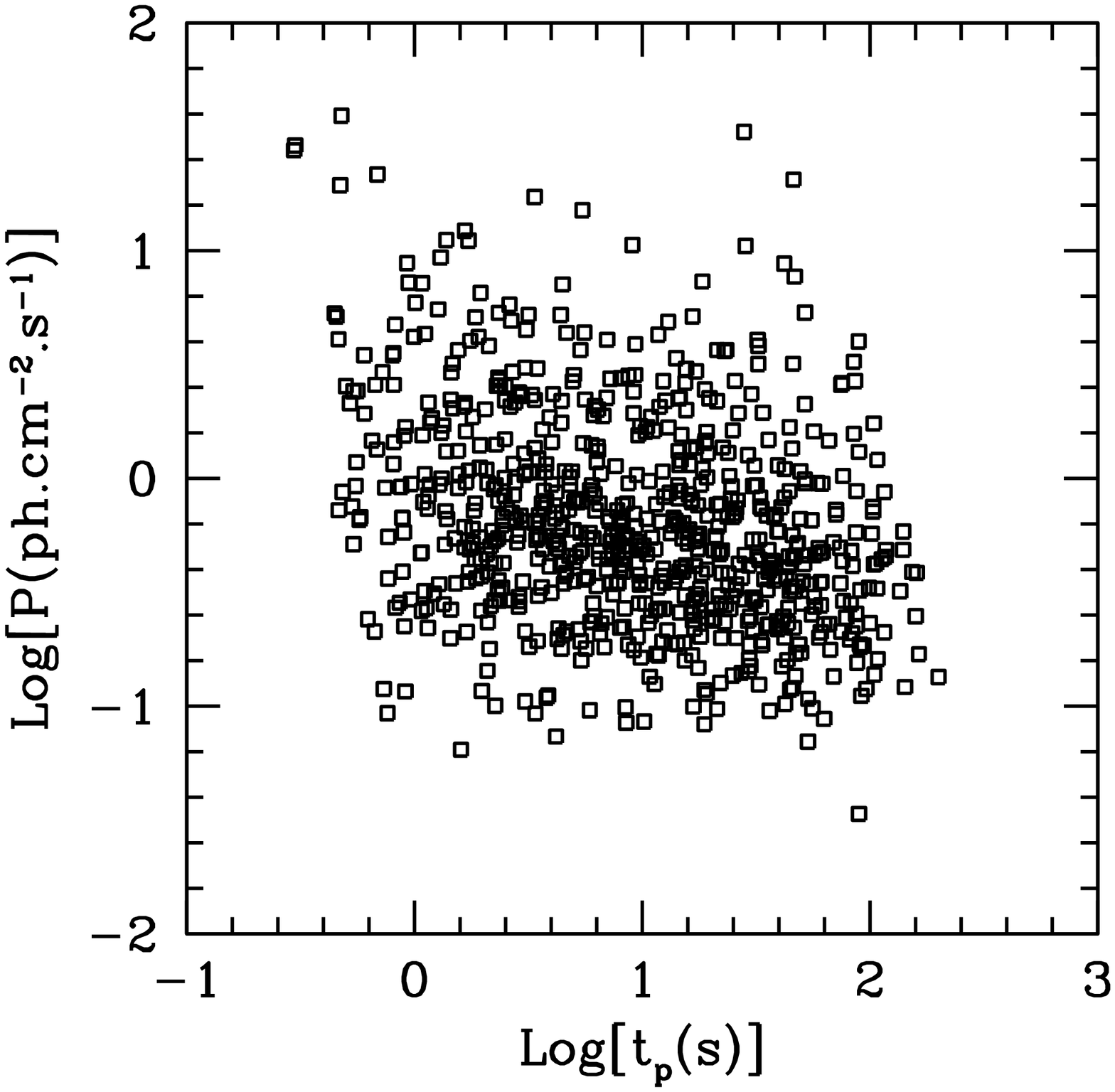}
\includegraphics{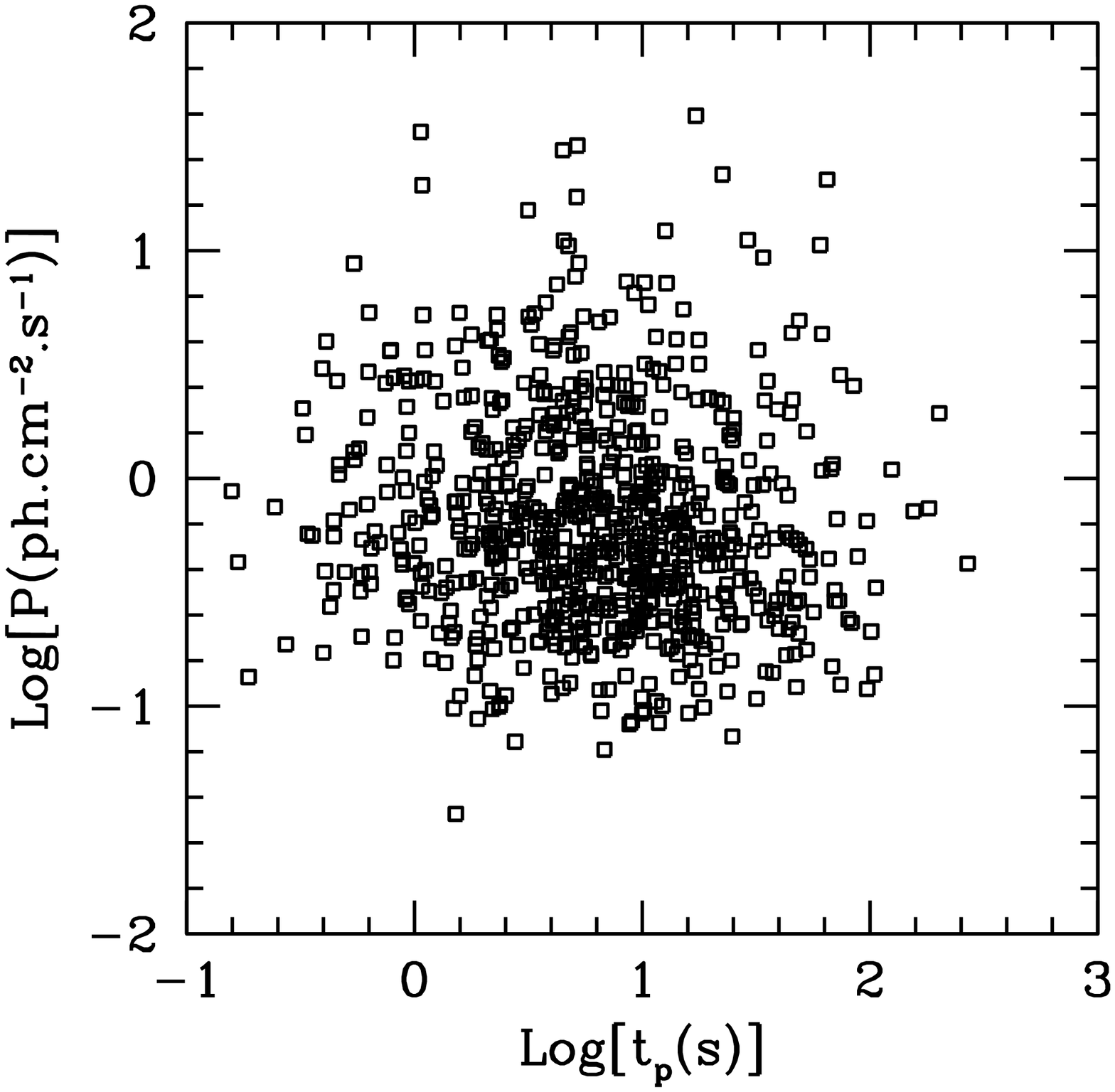}}
\end{center}
\caption{$t_{\rm p}^{\rm obs}$ -- $P$ diagram for a sample of 750 synthetic
 pulses corresponding to our reference case (see text for details). The
 left panel, which assumes the validity of the DLR looks very similar
to the data collected by Hakkila \& Cumbee (2009). The right
panel, with a log-normal distribution of pulse duration do not show a 
significant correlation between the two quantities.}
\end{figure*} 
\section{Discussion} 
Assuming the validity of the duration-luminosity relation,
we have tried to see if it can be understood in the context of the
internal shock model for the prompt emission of 
GRBs (Rees \& Meszaros, 1994). The isotropic
luminosity generated by internal shocks can be approximated by 
\begin{equation}
L=\dot E\, f(\kappa)\, \epsilon_e
\end{equation}
where $\dot E$ is the isotropic kinetic power in the relativistic flow,
$f(\kappa)$ is the efficiency of dissipation by internal shocks (which 
mainly depends on the contrast $\kappa$ between the maximum and minimum 
Lorentz factors) and $\epsilon_e$ is the fraction of the dissipated
energy which is transferred to electrons and eventually radiated.

It can be seen that the time scale $\tau$ for variability of the Lorentz
factor does not explicitely appear in Eq.(12). As the observed variability
of the prompt emission reflects that of the Lorentz factor in the
internal shock model (times the (1+$z$) dilation) any
duration-luminosity relation implies that $\tau$ should be in some way
linked to $\dot E$ and possibly also to $\kappa$ or $\epsilon_e$. One
could for example imagine that increasing $\dot E$ results in a more
unstable outflow where the Lorentz factor fluctuates on a shorter time
scale. This would induce a DLR which could become even more pronounced if
the amplitude of the fluctuations (and therefore $\kappa$)
also increases with $\dot E$. 
 
From an observational point of view there are some indications that 
$\dot E$ is anticorrelated with the opening angle of the relativistic
jet (Frail et al., 2001). A thinner jet drilling its way through the envelope of the
progenitor star could be more sensitive to Kelvin-Helmholtz
instabilities developing at its boundaries (Aloy et al., 2002). This would lead to a more
irregular outflow with a shorter time scale of variability of the
Lorentz factor, finally leading to a DLR. 
But clearly this discussion is somewhat speculative
and it remains that the internal shock model does not provide by itself
a simple and direct way to explain the DLR. 

Recently the internal shock model has also been criticized for a series of
reasons such as its low efficiency, the difficulty to explain 
the standard value $\alpha \sim -1$ of the low energy index 
of the spectrum (Ghisellini et al., 2000) or the possible complete suppression
of shocks if the flow is strongly magnetized.  
Proposed alternatives to internal shocks are reconnection processes 
(Giannos \& Spruit, 2007) 
relativistic turbulence (Narayan \& Kumar, 2009; Lazar, Nakar \& Piran, 2009) 
or comptonized photospheric emission
(Beloborodov, 2009). 
Unfortunately the modelling of these
mechanisms has not reached a degree of accuracy where detailed 
predictions can be made 
on the properties of pulses.  
\section{Conclusion}
We have considered the relations existing between spectral lags,
duration and luminosity in GRB pulses. Extending a previous work by
Hafizi \& Mochkovitch (2007) we have first shown that the lag over pulse
duration ratio does not vary much among pulses (remaining of the order
of a few percents). This result holds as long as the spectral softening
following pulse maximum is not limited to a decrease of the peak energy
but also affects the spectral indices, as indicated by the
observations. We have then included in our analysis the relation
between pulse duration and luminosity recently discovered by Hakkila et
al (2008). Combined to our results it allows to link all three quantities:
$\Delta t$, $t_{\rm p}$ and $L$. The lag-duration and lag-luminosity relations 
we obtain are in good agreement with the data. Also, they do not
strongly depend on the assumptions for the spectral parameters at pulse
maximum: values of $E_{\rm p}$, $\alpha$, $\beta$ and their derivatives,
Amati relation or log-normal distribution of
$E_{\rm p}$. 

Originally obtained with a limited set of only 12 pulses the DLR has recently
received further support from the analysis of another sample of 12 pulses
coming from 8 bursts detected by the HETE 2 satellite (Arimoto et al., 2010).
Its validity however still needs to be confirmed and we have therefore
proposed to test it in a different (statistical) way 
using the observational duration-peak photon flux 
($t_{\rm p}^{\rm obs}$ -- $P$) diagram. For that purpose, we have
adapted the Monte-Carlo code of Rossi, Daigne \& Mochkovitch (2006) to
generate a sample of synthetic pulses for which we predict the
observational $t_{\rm p}^{\rm obs}$ -- $P$ diagram. It appears that the
observed correlation $P\propto  t_{\rm p}^{-0.27}$ is reproduced only if
pulses satisfy the DLR. This conclusion remains valid when we vary the
pulse luminosity function and spectral properties ($E_{\rm p}$ obtained
from the Amati relation or having a log-normal
distribution). Nevertheless we cannot completely exclude some bias in
the pulse selection and characterization process which could contribute
to the observed relation even in the absence of a DLR.

We have finally confronted the DLR to the prediction of the internal
shock model for the prompt emission. It appears that the DLR cannot be
obtained as a direct and simple consequence of the 
model. Additional assumptions are required, for example the possibility
that the relativistic outflow becomes more unstable and variable 
when the injected kinetic power increases. Proposed alternatives to internal
shocks -- reconnection, relativistic turbulence, 
comptonized photosphere -- 
still don't have the
predictive power to test if they can explain the DLR. 

In a future development of this work we plan to extend our analysis 
of the pulse properties (width and spectral lags) to other energy
ranges. Data are sparse in the optical but the detailed light curve of
the ``naked-eye burst'' GRB 080319b has for example revealed interesting 
correlations between spectral lags at high and low energy
(Stamatikos et al., 2009). At very high energy, Fermi observations have
shown delays in the onset of the LAT component with respect to the MeV
emission. Understanding the origin of these behaviors will provide clues for
a better understanding of the prompt emission of GRBs.
 
\begin{acknowledgements}
It is a pleasure to thank Jon Hakkila for his numerous advices and for
having sent to us unpublished materials and data. We also thank Makoto
Arimoto who has kindly answered our questions about the HETE 2 data.  

\end{acknowledgements}

  \end{document}